\documentclass[aps,prl,twocolumn,superscriptaddress,footinbib]{revtex4-2}
\usepackage{graphicx}  % Include figure files
\usepackage{subfigure}
\usepackage{dcolumn} % Align table columns on decimal point
\usepackage{bm}% bold math
\usepackage{amssymb}   % for math
\usepackage{amsfonts}   % for math
\usepackage{comment}
\usepackage[colorlinks,linkcolor=blue,anchorcolor=blue,citecolor=blue,urlcolor=blue]{hyperref}
\usepackage{xcolor}
\usepackage{soul}
\usepackage{tabularx,booktabs,ragged2e,multirow}
\newcolumntype{C}{>{\centering\arraybackslash}X}
\usepackage{multirow}
\usepackage{amsmath}
\usepackage{mathrsfs}
\usepackage{mathcomp}
\usepackage{textcomp}
\usepackage{dsfont}
\usepackage{esint}
\usepackage{braket}
\usepackage{textgreek}
\usepackage{lipsum}
\usepackage{marvosym} 
\usepackage{centernot}
\usepackage{cancel}
\usepackage{dashrule}
\begin{document}
\preprint{APS/123-QED}

%\title{Low-energy signature and magnetic suppression of NH skin effects}
\title{Magnetic suppression of non-Hermitian skin effects}

\author{Ming Lu}
\affiliation{Beijing Academy of Quantum Information Sciences, Beijing 100193, China}
\affiliation{International Center for Quantum Materials, School of Physics, Peking University, Beijing 100871, China}
\author{Xiao-Xiao Zhang}
\email{Correspondence: xiaoxiao.zhang@riken.jp}
\affiliation{Department of Physics and Astronomy \& Stewart Blusson Quantum Matter Institute, University of British Columbia, Vancouver, BC, V6T 1Z4 Canada}
\affiliation{RIKEN Center for Emergent Matter Science (CEMS), Wako, Saitama 351-0198, Japan}

\author{Marcel Franz}
\affiliation{Department of Physics and Astronomy \& Stewart Blusson Quantum Matter Institute, University of British Columbia, Vancouver, BC, V6T 1Z4 Canada}

%\date{\today}

\newcommand{\br}{{\bm r}}
\newcommand{\bk}{{\bm k}}
\newcommand{\bq}{{\bm q}}
\newcommand{\bp}{{\bm p}}
\newcommand{\bv}{{\bm v}}
\newcommand{\bmm}{{\bm m}}
\newcommand{\bA}{{\bm A}}
\newcommand{\bE}{{\bm E}}
\newcommand{\bB}{{\bm B}}
\newcommand{\bH}{{\bm H}}
\newcommand{\bd}{{\bm d}}
\newcommand{\bzero}{{\bm 0}}
\newcommand{\bOmega}{{\bm \Omega}}
\newcommand{\bsigma}{{\bm \sigma}}
\newcommand{\bJ}{{\bm J}}
\newcommand{\bL}{{\bm L}}
\newcommand{\bS}{{\bm S}}
\newcommand\dd{\mathrm{d}}
\newcommand\ii{\mathrm{i}}
\newcommand\ee{\mathrm{e}}
\newcommand\zz{\mathtt{z}}
\makeatletter
\let\newtitle\@title
\let\newauthor\@author
\def\ExtendSymbol#1#2#3#4#5{\ext@arrow 0099{\arrowfill@#1#2#3}{#4}{#5}}
\newcommand\LongEqual[2][]{\ExtendSymbol{=}{=}{=}{#1}{#2}}
\newcommand\LongArrow[2][]{\ExtendSymbol{-}{-}{\rightarrow}{#1}{#2}}
\newcommand{\cev}[1]{\reflectbox{\ensuremath{\vec{\reflectbox{\ensuremath{#1}}}}}}
\newcommand{\red}[1]{\textcolor{red}{#1}} %for displaying red texts
\newcommand{\blue}[1]{\textcolor{blue}{#1}} %for displaying blue texts
\newcommand{\green}[1]{\textcolor{orange}{#1}} %for displaying blue texts
\newcommand{\mytitle}[1]{\textcolor{orange}{\textit{#1}}}
\newcommand{\mycomment}[1]{} %for commenting out
\newcommand{\note}[1]{ \textbf{\color{blue}#1}}
\newcommand{\warn}[1]{ \textbf{\color{red}#1}}

\makeatother

\begin{abstract}
Skin effect, where macroscopically many bulk states are aggregated towards the system boundary, is one of the most important and distinguishing phenomena in non-Hermitian quantum systems. We discuss a new aspect of this effect whereby, despite its topological origin, applying magnetic field can largely suppress it. Skin states are pushed back into the bulk and the skin topological area, which we define, is sharply reduced. As seen from exact solutions of representative models this is fundamentally rooted in the fact that the applied magnetic field restores the validity of the low-energy description that is rendered inapplicable in the presence of non-Bloch skin states. We further study this phenomenon using rational gauge fluxes, which reveals a unique irrelevance of the generalized Brillouin zone in the standard non-Bloch band theory of non-Hermitian systems. 
\end{abstract}
%\pacs{71.10.Pm, 71.27.+a, 72.15.Nj, 72.15.Rn}
\keywords{}

\maketitle
% \tableofcontents
% \newpage
% \clearpage

%\section{Introduction}
\mytitle{Introduction}.--
Hermicity has long been accepted as a basic requirement in quantum mechanics \cite{Dirac1942}. However, it was discovered that both reality of energy spectra and unitarity of time evolution can hold in the parity-time ($\mathcal{PT}$) symmetric non-Hermitian (NH) Hamiltonians \cite{Bender1998, Bender2002, Bender2007}. 
This finding and its experimental realizations \cite{Guo2009,Rueter2010,Schindler2011,Chtchelkatchev2012,Bittner2012, Regensburger2012, Bender2013, Hang2013, Regensburger2013, Feng2012, Peng2014, Chang2014, Hodaei2014, Fleury2015, Zhang2016, Weimann2016, Assawaworrarit2017, Xiao2017, Liu2018,Zhang2018a,Wu2019c, Klauck2019, Shao2020, Assawaworrarit2020, Wang2020} focused interest on the properties of NH systems \cite{Moiseyev2009, Mostafazadeh2010, Feng2017, ElGanainy2018, Alvarez2018, Ashida2020}. As an effective description, NH Hamiltonians have a wide range of applications, including systems with energy or matter source and drains \cite{ElGanainy2007,Musslimani2008, Makris2008, Klaiman2008, Single2014, Ezawa2019b, Ezawa2019c, Hofmann2019, Stegmaier2021}, modeling of subsystems with an environment \cite{Hatano1996,Rotter2009, Avila2019,Bergholtz2019,Yang2021}, and quasiparticles in solids with interaction or disorder \cite{Liu2014, Kozii2017, Zyuzin2018, Yoshida2018, Yamamoto2019, Zyuzin2019,Moors2019, Papaj2019}. Unique features have been found with no Hermitian counterparts including exceptional points and rings \cite{Dembowski2001,Dembowski2004, Berry2004, Heiss2012, Wiersig2014, Chen2017, Hodaei2017, Zhang2019e, Miri2019, Wiersig2020, Wiersig2020a, Bergholtz2021, Zhen2015, Xu2017, Cerjan2018, Cerjan2019}, extended topological classifications \cite{Shen2018, Gong2018, Kawabata2019, Torres2019, Ghatak2019}, and NH skin effect (NHSE) under open boundary condition (OBC) \cite{Alvarez2018a, Yao2018a, Xiong2018, Lee2019, Ezawa2019, Ezawa2019a, Brandenbourger2019, Ghatak2020, Hofmann2020, Xiao2020, Budich2020, Helbig2020, Weidemann2020, Zhang2021}. In NHSE states are localized near the boundaries, which leads to the breakdown of conventional bulk-boundary correspondence and motivates the concept of generalized Brillouin zone (GBZ) \cite{Lee2016, Yao2018,  Kunst2018, Yokomizo2019, Kawabata2020, Yang2020a}.
%\mycomment{For instances, both eigenvalues and eigenvectors coalesce at the NH exceptional points\cite{Dembowski2001,Dembowski2004, Berry2004, Heiss2012}, around which the degeneracy is very vulnerable to perturbations and can be exploited for ultra-sensitive sensors\cite{ Wiersig2014, Chen2017, Hodaei2017,Zhang2019e, Miri2019, Wiersig2020, Bergholtz2021}. Another distinct feature is the NH skin effect (NHSE), where a finite fraction of eigenstates are localized at the boundary\cite{Alvarez2018a, Yao2018a, Xiong2018, Lee2019, Hofmann2019a, Brandenbourger2019, Ghatak2020, Xiao2020, Helbig2020, Weidemann2020, Zhang2021}. It directly leads to the breakdown of conventional bulk boundary correspondence and motivate the concept of generalized Brillouin zone (GBZ) \cite{Lee2016, Yao2018,  Kunst2018, Yokomizo2019, Kawabata2020, Yang2020a}.} 
Intriguingly, NHSE has an unconventional topological origin in terms of complex energies instead of wavefunctions \cite{Okuma2020, Zhang2020}. Being a robust and ubiquitous phenomenon in NH systems, an interesting question arises:  Is it possible to control or manipulate NHSE? Given its nonperturbative effect in Hermitian physics one might suspect that magnetic field (real or synthetic) could be used here. 
Techniques to generate (synthetic) magnetic field have been well-established in candidate NH systems, such as cold atoms \cite{Madison2000, AboShaeer2001, Lin2009}, photonic and acoustic structures \cite{Rechtsman2012, Li2014a, Lumer2019, Wen2019}, electric circuits \cite{Ningyuan2015,Albert2015,Zhao2018,XXZ:nHcircuit}, and conventional solid-state systems \cite{Guinea2009, Levy2010, Nigge2019}. Yet its effects in NH systems have only been addressed in some specific scenarios \cite{Shen2018a, Okuma2019,XXZ:nHcircuit}. %\st{In fact, techniques to generate (synthetic) magnetic field have been well established in several kinds of potential NH systems, such as cold atoms systems\cite{Madison2000, AboShaeer2001, Lin2009}, photonic and acoustic systems\cite{Rechtsman2012, Li2014a, Lumer2019, Wen2019}, and the conventional solid states systems\cite{Guinea2009, Levy2010, Nigge2019}.}

\begin{figure}[t]
\includegraphics[width=8.6cm]{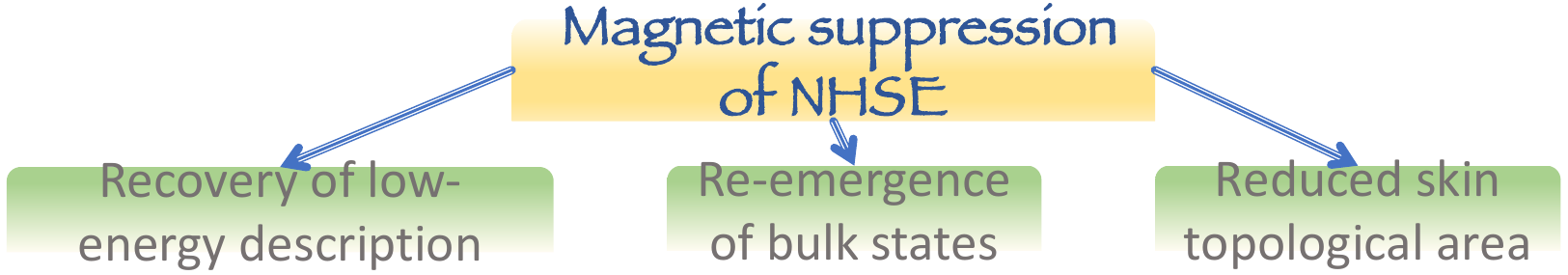}
\caption{Manifestations of the magnetic suppression of NHSE.}\label{Fig:cartoon}
\end{figure}

Here, we study a more generic problem, namely the interplay of magnetic field and NHSE. Although the latter is topological as discussed in more detail below, we find that magnetic field can strongly suppress it. Our key findings are summarized in Fig.~\ref{Fig:cartoon}. The skin topology, being a global property of the lattice model, leads to macroscopically many skin states. As we show this invalidates the low-energy description that carries only local information. However, the skin states localized at the boundary can be pushed back to the bulk by the applied magnetic field. As elucidated by analytical calculations in representative models, this phenomenon originates from the nonperturbative nature of magnetic field: the fundamental breakdown of low-energy description can be recovered. As a quantitative indicator, the skin topological area, defined as the regions in the complex-energy plane where the skin winding number is nontrivial, is sharply reduced in corresponding lattice models with irrational and rational gauge fluxes. Specifically, rational gauge-flux calculation enables us to show an unexpected irrelevance of GBZ to the suppression of NHSE, contrary to the intuition that suppression results from deformation of GBZ towards a conventional Brillouin zone.

%The skin topology as a global property of the lattice system leads to macroscopically many skin states. This NHSE disables the low-energy description with only local information; an additional magnetic field completely recovers the low-energy theory by capturing the newly formed bulk states.
%(b) The possibly surjective mapping from NH lattice systems to low-energy models fundamentally forbids identifying NHSE from low-energy models, i.e., the stop-sign in (a).
%\warn{In (c), several options: add 'continuum skin bands acquire discreteness'; replace the 2nd one by 'Re-emergence of bulk discrete levels' or 'Bulk discrete spectrum emerges from continuum (skin) bands'? .}
%AA model\cite{Jiang2019}

% \section{Model systems}
% \subsection{2-band model}
 
% \subsection{1-band model}

%\section{Model systems with unidirectional skin effect}
\mytitle{Unidirectional skin systems}.--
We consider a one-band model
\begin{align}
    H = 2 t_x\cos{k_x} + 2 t_y \cos{k_y} + 2\ii\gamma_y\sin{k_y} \label{eq:H_1band1}
    % H = (\kappa+t\cos{k_x})\cos{k_y} + \ii\gamma_y\sin{k_y} \label{eq:H_1band2}
\end{align}
and a two-band model $H=\bd\cdot\bsigma$ with
\begin{equation}\label{eq:H_2band}
%\begin{split}
 d_x = \ii\gamma_x - \kappa +t_y\cos{k_y} - t_x\cos{k_x},
 d_y = t_y\sin{k_y} + \ii\gamma_y
%\end{split} 
\end{equation}
and $d_z = \Delta$. Lattice constant is set to unity henceforth.
In the Hermitian case when $\gamma_{x,y}=0$, Eq.~\eqref{eq:H_1band1} has the band minimum at $\bk =(\pi,\pi)$; Eq.~\eqref{eq:H_2band} has two Dirac points ($\Delta=0$) or two band edges separated by a gap ($\Delta>0$) when $\kappa<t_x+t_y$ and is otherwise gapped with the band edges at $\bk =(\pi,0)$ when $\kappa>t_x+t_y$. %while Eq.~\eqref{eq:H_1band2} has the band minimum at $\bk =(0,\pi)$.
%\subsection{Unidirectional topological skin winding number}
These two models share one unique property: the NHSE is present \textit{if and only if} $y$-direction is open and $\gamma_y\neq0$. This can be directly seen from evaluating the skin topological winding number \cite{Okuma2020,Zhang2020}
\begin{equation}\label{eq: winding_num}
   w(E_0) =\frac{1}{2\pi\ii}\int_0^{2\pi} \dd k \frac{\dd}{\dd k} \log \det[H(k)-E_0],
\end{equation}
where $k$ %representing either $k_x$ or $k_y$ 
is the conventional Bloch momentum in any open direction under question \textit{as if} it has a periodic boundary condition (PBC) and a Brillouin zone. For $E_0$ outside the spectrum of $H(k)$ this extends the definition of a gap to the complex-energy plane $\mathbb{C}$. Eq.~\eqref{eq: winding_num} actually manifests the unique point-gap topology that is entirely distinct from Hermitian systems and those NH systems continuously deformable to them \cite{Kawabata2019a,Okuma2020}.
The necessary and sufficient condition for NHSE along that direction under question is the existence of $E_0\in\mathbb{C}$ such that $w(E_0)\neq0$.
It can be nonzero only for the foregoing condition, otherwise the trajectory of $\det[H(k)-E_0]$ collapses into a retracing arc without interior. 

Such a unidirectional feature of NHSE is crucial to investigating the system under magnetic field. 
We consider a uniform magnetic field along $z$-direction for our two-dimensional system, which is conveniently generated by the minimal coupling $\Pi_i\rightarrow k_i-A_i$. To use Eq.~\eqref{eq: winding_num} with a conserved $k_y$ to study NHSE in $y$-direction for instance, we can only use vector potential $\bA =B(0,x)$ and are thus forced to leave $x$-direction open. As the skin phenomenon can have significant impact on the physical properties\,\cite{Okuma2021}, we would like $x$-direction with OBC to be conventional such that no qualitative change is introduced; therefore, the unidirectional NHSE along $y$-direction only is the simplest and clearest choice. Although we make use of a specific gauge we checked that the observable  consequences %of the skin effect 
(e.g.\ the  wavefunction localization near the edge) are gauge invariant as physically expected.  

%\section{Failure and recovery of low-energy description}
\mytitle{Failure and recovery of low-energy description}.--
In the following, we inspect the low-energy effective models around the band edges of the lattice models, which turn out to fall into three major classes exhausting the common possibilities: (a) nonrelativistic Schr\"odinger-type from Eq.~\eqref{eq:H_1band1} around $\bk =(\pi,\pi)$: $h_a = -m + t_x \Pi_x^2  + t_y (\Pi_y - \ii\frac{\gamma_y}{t_y})^2$ with $m=2t_x+2t_y-\frac{\gamma_y^2}{t_y}$; 
(b) relativistic Dirac-type from Eq.~\eqref{eq:H_2band} around $\bk =(\frac{\pi}{2},0)$ when $\kappa=t_y$: $h_b=\sum_{i=x,y}{(t_i\Pi_i+\ii\gamma_i)\sigma_i}+\Delta\sigma_z$; 
(c) a mixture of both from Eq.~\eqref{eq:H_2band} around $\bk =(\pi,0)$ when $\kappa>t_x+t_y$: $h_c=[-\mathfrak{m}-\frac{1}{2}t_x\Pi_x^2]\sigma_x +(t_y\Pi_y+\ii\gamma_y)\sigma_y +\Delta\sigma_z$ with $\mathfrak{m}=\kappa-t_y-t_x-\ii\gamma_x$. 
We henceforth use fraktur letters %like $\mathfrak{m}$ 
to mark certain complex quantities. %and $\bk $ in low-energy models means the deviation from the band minimum.
Our considerations lead to the general relation: While the NHSE impedes the low-energy effective description (where skin states are essentially missing and only trivial Bloch solutions exist),  adding an external magnetic field, which is a \textit{nonperturbative} effect, in general recovers the validity of low-energy models.

We exemplify with the representative case (b), which can be reduced to a single equation for the second wavefunction component (setting $t_{x,y}=1$ for brevity)
\begin{equation}\label{eq:Dirac_2ndorder_shift_main}
    \psi_2'' - 2\gamma_y \psi_2' + [\tilde{\varepsilon}^2 -  B^2(y-\mathfrak{y}_0)^2] \psi_2=0
\end{equation}
where $\tilde{\varepsilon}^2=\varepsilon^2 - \Delta^2 + \gamma_y^2 - B,\mathfrak{y}_0=-(k_x+\ii\gamma_x)/B$.
The nonrelativistic type $h_a$ can be exactly mapped to Eq.~\eqref{eq:Dirac_2ndorder_shift_main} because it combines the two linear dispersion components in $h_b$ to effectively get a Schr\"odinger-like one. The mixed type $h_c$ requires a more involved analysis due to non-Hermiticity and inhomogeneous dispersion, given in Supplemental Material (SM)\cite{SM}. Crucially, all three cases exhibit similar and closely related physical behavior, justifying the generality of our conclusions.% as summarized in Fig.~\ref{Fig:cartoon}(b).
\begin{figure}[t]
\includegraphics[width=8.6cm]{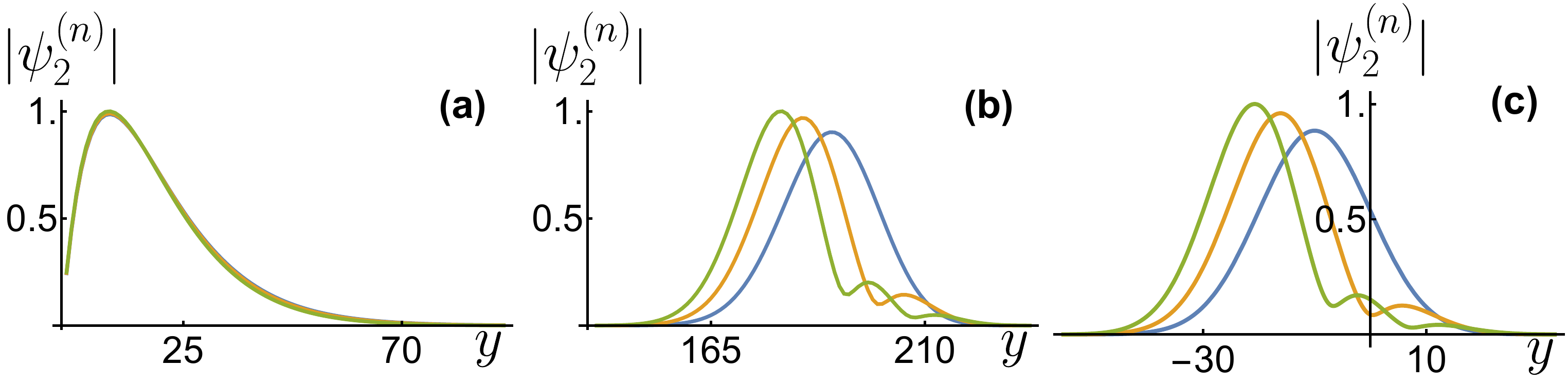}
\caption{Amplitude $|\psi_2^{(n)}|$ of lowest energy states (blue, orange, and green respectively for $n=0,1,2$) from the second wavefunction component of the representative Dirac-type Hamiltonian. 
The Schr\"odinger and mixed types have similar features and are shown in SM. Lattice calculations (a,b) with $N=400$ sites along $y$-direction assume $y=0$ as the left boundary: (a) skin states aggregate towards the left boundary and do not have low-energy description when $B=0$; (b) finite magnetic field $B=0.01$ moves these states back to the bulk. (c) Low-energy exact solutions from Eq.~\eqref{eq:Dirac_2ndorder_shift_main} defined along $y\in(-\infty,\infty)$ are recovered by $B$ and match with panel (b). Parameters $\gamma_x=0.02,\gamma_y=-0.1,\kappa=t_x=t_y=1$.}\label{Fig:wave_functions}
\end{figure}
%Spatial coordinate $y$ is in units of lattice constant set to unity. 

When $B=0$, Eq.~\eqref{eq:Dirac_2ndorder_shift_main} reduces to the form $\psi_2''-2\gamma_y\psi_2'+(\mathcal{K}(\varepsilon)^2+\gamma_y^2)\psi_2=0$ where $\mathcal{K}(\varepsilon)^2=\varepsilon^2-\Delta^2-(k_x+\ii\gamma_x)^2$.
Choosing a complex $\mathcal{K}$, the general solution $\psi_2(y)=\ee^{\gamma_yy}(A\sin{\mathcal{K}y}+B\cos{\mathcal{K}y})$ can accommodate the trivial Bloch state as a bounded solution with ${\rm Re}\mathcal{K}$ the Bloch momentum. This is to be contrasted with the numerical solution of the lattice model shown in Fig.~\ref{Fig:wave_functions}(a) that exhibits NHSE.
One might try imposing, say, $\psi_2(y<0)\equiv0$, in order to describe skin states accumulated near the left boundary at $y=0$.
However, this fails to provide a valid low-energy description comparable to conventional Hermitian edge states for at least two reasons:
(i) The actual decay rate of wavefunction cannot be determined \textit{within} the low-energy model itself, e.g., through a boundary condition to specify $\mathcal{K}$. Such information is missing from the low-energy model since it is given by the continuum of generally state-dependent GBZ radius $|\beta|$, defined for the lattice system and related to the skin topology as discussed later.
In contrast, the Hermitian low-energy theory is in general self-contained and able to fully characterize the edge states. The macroscopically large number of skin states, on the other hand, manifests itself as a \textit{nonlocal} property arising from the unconventional topology.
(ii) We can see the failure in a more self-evident way. The difference between global topology and local description %, indicated in Fig.~\ref{Fig:cartoon}(b), 
leads to a caveat: the correspondence between NH systems and characteristic low-energy models is \textit{surjective}. As an extreme example, due to lattice symmetry, $x,y$ are on the same footing in the foregoing low-energy model $h_b$, i.e., the unidirectional topological skin information is completely \textit{concealed}. Note that, including higher-order terms in the low-energy theory, which is a nonuniversal procedure, cannot guarantee to remedy these problems in general. Therefore, the failure of the low-energy description is an inevitable feature of NHSE systems, deeply rooted in its inability to capture the full nonlocal topological information.

The situation is, however, entirely different when $B\neq0$. Eq.~\eqref{eq:Dirac_2ndorder_shift_main} now admits a well-defined solution
\begin{equation}\label{eq:phi_n_main}
\psi_2^{(n)}
\mycomment{=(\sqrt{\pi}l_B n! 2^{n})^{-\frac{1}{2}}  \ee^{-y^2/2l_B^2 -(k_x+\ii\gamma_x-\gamma_y)y} H_n((y+\frac{k_x+\ii\gamma_x}{b})/l_B) }
\propto \mycomment{(\sqrt{\pi}l_B n! 2^{n})^{-\frac{1}{2}}}  \ee^{-(y-\mathfrak{y}_0)^2/2l_B^2+\gamma_yy} \mathcal{H}_n((y-\mathfrak{y}_0)/l_B)
\end{equation} 
with magnetic confinement length $l_B=B^{-1/2}$ and $\mathcal{H}_n$ the $n$th \textit{Hermite polynomial} holomorphic in $\mathbb{C}$. %and
% \begin{equation}\label{eq:HeunGenSol_main}
%     \psi_2(\mathsf{y})= \sum_{s=\pm} C_s\, \ee^{(\gamma_y-s\mathfrak{m})\mathsf{y} - z_s^3/2} \mathscr{H}_\mathrm{T}(\alpha,\beta_s,\gamma,z_s)
% \end{equation}
% with integration constants $C_\pm$, $r=(\frac{3}{B^2})^{\frac{1}{3}}$,  $\alpha=r^2(\tilde\varepsilon^2- \gamma_y^2),\beta_\pm=\pm3,\gamma=2r\mathfrak{m},z_\pm=\pm\mathsf{y}/r$ and $\mathscr{H}_\mathrm{T}$ the \textit{triconfluent Heun function} \cite{Ronveaux1995,Slavyanov2000}. Note that Eq.~\eqref{eq:mix_type_main} is solved not at the conventional polynomially truncated subspace; hence the irregular singularity at $\infty$ complicates the analysis and the quantum number $n$ is implicit in Eq.~\eqref{eq:HeunGenSol_main}. 
The mixed type $h_c$ assumes a more complex solution involving the \textit{triconfluent Heun function} $\mathscr{H}_\mathrm{T}(\alpha,\beta,\gamma,z)$ \cite{SM,Ronveaux1995,Slavyanov2000}. %and is solved not at the conventional polynomially truncated subspace\cite{SM,Ronveaux1995,Slavyanov2000}.
Remarkably, all three low-energy models share a few key features highlighted by Eq.~\eqref{eq:phi_n_main}, as confirmed by lattice-model and exact solutions shown in Fig.~\ref{Fig:wave_functions}(b,c). We observe that (i) Solutions are now \textit{bulk} states confined by $B$ with a discrete spectrum bounded from below; the wavefunctions are otherwise divergent due to non-Hermiticity as $B\rightarrow0$, indicating the nonperturbative nature. 
(ii) Wavefunctions are \textit{nodeless} if and only if $\gamma_x\neq0$ since the quasiparticle now effectively moves in $\mathbb{C}$ due to the $\gamma_x/B$ shift along imaginary $y$-direction easily seen from Eq.~\eqref{eq:phi_n_main}. This is absent in the one-band model Eq.~\eqref{eq:H_1band1} or $h_a$. (iii) The apparent imaginary momentum shift $\gamma_y$ in $\Pi_y$, seen from its appearance in all low-energy Hamiltonians, adds a slanting asymmetry to the amplitude or peak height along $y$. One important distinction lies in that the two wavefunction components $\psi_{1,2}$ for the mixed type $h_c$ share the \textit{same} quantum number $n$, i.e., the same number of amplitude peaks/nodes, in stark contrast to the Dirac-type with $\psi_1^{(n+1)}$ in pair with Eq.~\eqref{eq:phi_n_main}.

The ability to recover a valid low-energy description is physically due to an exceptionally robust magnetic confinement effect that counteracts non-Hermiticity and works for various dispersions. The wavefunction is, despite NHSE, forced to localize in the bulk within a confinement length scale $l_B$. %In a small system, the system size $L\lesssim l_B$ competing with the magnetic confinement can complicate the probability distribution. 
For a macroscopic system of our main interest, finite $l_B$ controls the possible NH divergence in the wavefunction and macroscopically many skin states, if not all, are no longer destined to be pushed all the way to the edges. The skin wavefunctions are  transformed back to normal bulk states that admit valid low-energy description. When $l_B$ is comparable to or longer than a mesoscopic/microscopic system, the size constraint will otherwise become relevant and reduce the nonperturbative effects.

\mytitle{Magnetic suppression of NH skin effect}.--
To visualize the NHSE and its magnetic suppression, we need to open $y$-direction and use the gauge $\bA =B(-y,0)$ with $k_x$ a good quantum number. We thus obtain effective one-dimensional lattice models $H(k_x)$ \cite{SM}. %\red{Eqs. (\ref{eq:finite-size-openY-1band}) and (\ref{eq:finite-size-openY-2band})}. 
As shown in Fig.~\ref{Fig:wave_functions}, without magnetic field,  wavefunctions are localized near the left edge due to NHSE. Applying the magnetic field, NHSE is suppressed and wavefunctions move away from the edges towards the bulk. It can be observed that the low-energy states in the lattice models under magnetic field are well captured by the low-energy continuum model. This magnetic suppression also holds for states at higher energy as a much more extensive global property; we find that the average center of almost all wavefunctions moves towards the bulk interior\cite{SM}.

\begin{figure}[t]
\includegraphics[width=8.6cm]{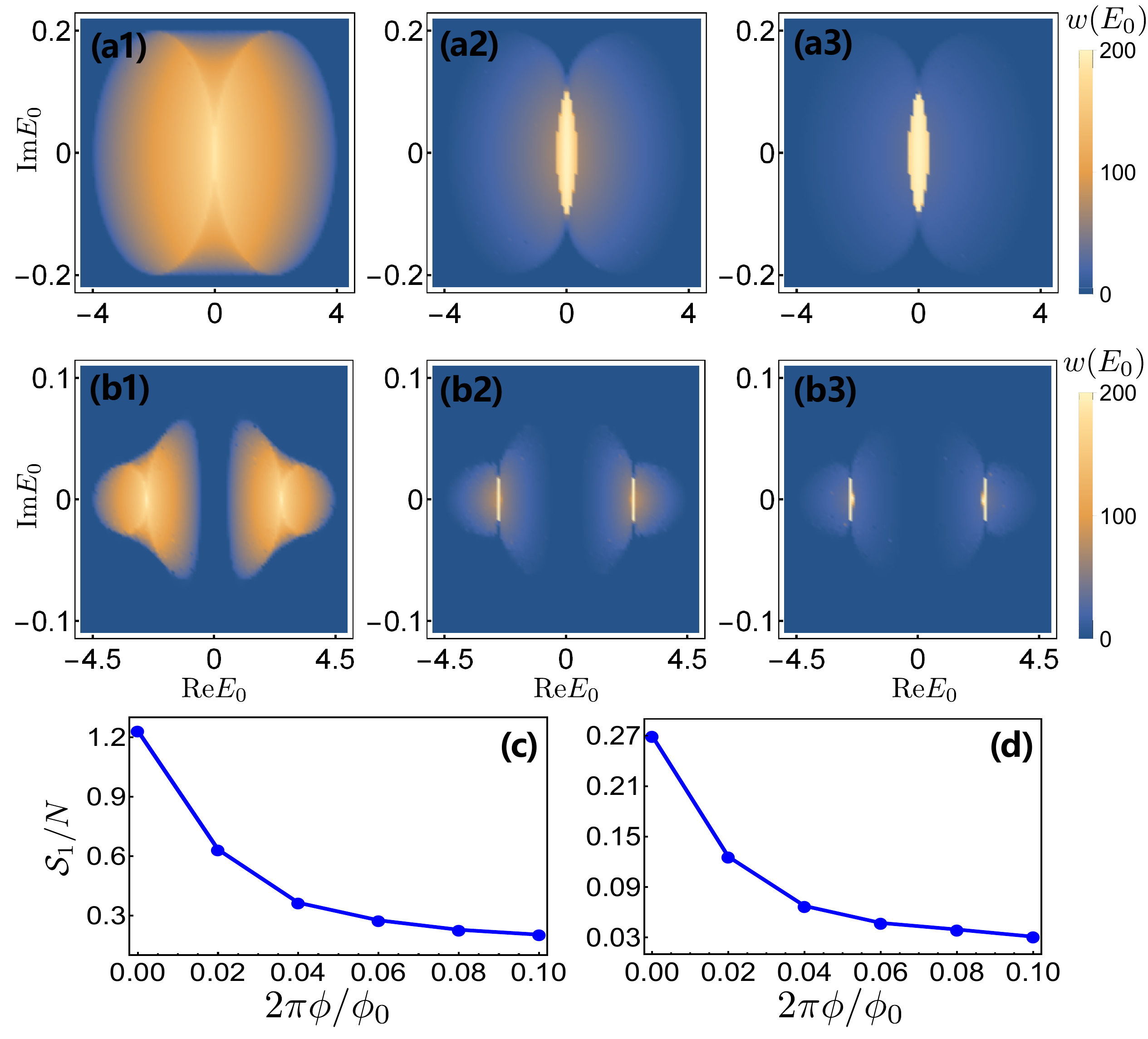}
\caption{Color-coded winding numbers in the complex-energy plane for (a1-a3) one- and (b1-b3) two-band lattice models with magnetic flux $2\pi\phi/\phi_0=0, 0.04, 0.08$, respectively. Skin topological area $\mathcal{S}_1/N$ as a function of magnetic flux for (c) one- and (d) two- band models. Lattice sites $N=200$, $\gamma_y=0.1$, $t_x=t_y=1$ and $\gamma_x=0,\kappa=2.5$.}\label{Fig:Landau_skin}
\end{figure}

This phenomenon also has a topological aspect closely related to Eq.~\eqref{eq: winding_num}. From this perspective, the presence of skin modes corresponds to the existence of $E_0$ in $\mathbb{C}$ that makes $w(E_0)$ nonzero. The set of all such points forms a special complex-energy plane region, which is not necessarily simply connected [e.g., Fig.~\ref{Fig:Landau_skin}(b)]. We propose that, the winding number-weighted area of this nontrivial region, which we henceforth call \textit{skin topological area}, can serve as a quantitative indicator of the suppression phenomenon. This also places the topological number $w(E_0)$, which lacks a simple interpretation in terms of protected edge states as in the Hermitian case, in the concrete physical context of the skin effect strength. We will show below that applying magnetic field causes significant reduction in the skin topological area.
% is open. The obtained effective one dimensional lattice models $H(k_y)$, described by Eqs. (\ref{eq:finite-size-openX-1band}) and (\ref{eq:finite-size-openX-1band}), are then used in Eq. (\ref{eq: winding_num}).

%We choose $N=100$ unit cells to calculate the winding numbers for both models. 
The calculation is performed in the gauge $\bA = B(0, x)$ with the system open along $x$-direction as previously discussed.
Fig.~\ref{Fig:Landau_skin} illustrates how the skin area $\mathcal{S}_1$ shrinks as the magnetic field increases. For $B=0$ in particular, $\mathcal{S}_1$ for the one-band model Eq.~\eqref{eq:H_1band1} is approximately $4\pi\gamma_y t_y N$ with $N$ the number of unit cells in the $x$-direction. This can be understood by noting that $x$-direction has no NHSE and hence OBC is effectively the same as PBC. The winding of $H$ at any fixed $k_x$ as $k_y$ varies from $0$ to $2\pi$ encloses an ellipse in $\mathbb{C}$ with semi-axes $2t_y,2\gamma_y$ and hence the area $4\pi\gamma_y t_y$. Similarly, for the two-band model, the skin topological area $\mathcal{S}_1$ when $B=0$ is equal to $\frac{1}{2\pi}\int_0^{2\pi} \mathcal{S}_1(k_x)\mathrm{d}k_x$, where $\mathcal{S}_1(k_x)$ is the skin area from Eq.~\eqref{eq:H_2band} at a given $k_x$. $\mathcal{S}_1$ is obviously an extensive quantity in the orthogonal direction without NHSE and we thus plot $\mathcal{S}_1/N$ in Fig.~\ref{Fig:Landau_skin}(c,d).

\mytitle{Rational magnetic flux and irrelevance of GBZ}.--
An alternative but informative viewpoint on the phenomena discussed so far is placing the NH system under magnetic field with a rational gauge flux $B=p\phi_0/q$, where $p,q$ are coprime integers and $\phi_0$ denotes the flux quantum.  For an $m$-band model in the gauge $\bA = B(0, x)$ we have a $mq\times mq$ Hamiltonian $H(\tilde{k}_x, k_y)$ with $\tilde{k}_x \in [0,2\pi/q),k_y\in [0,2\pi)$. Rational flux implies periodicity with an enlarged unit cell which confers a degree of analytic tractability in terms of the GBZ in the non-Bloch band theory.  We employ this formalism in the calculation of skin topological area. 
The GBZ predicts the band spectrum under OBC in the macroscopic limit \cite{Yao2018a,Yokomizo2019,Kawabata2020}. Deviation of a GBZ $\beta(\theta)$, parametrized by a generic $\theta\in[0,2\pi)$, from the conventional unit-circular Brillouin zone in $\mathbb{C}$, $\ee^{\ii k}$ for $k\in[0,2\pi)$, can imply the NHSE and determine the state-dependent exponential decay rate $\beta$.

With the non-Bloch momentum substitution $\ee^{\ii k_y}\rightarrow\beta$, we find that 
the determinant in Eq.~\eqref{eq: winding_num} takes the form
\begin{equation}\label{eq:Det}
\det[ H(\beta)-E_0 ]= A_0 + A_+ \beta^q + A_- \beta^{-q}
\end{equation}
where $E_0$ enters only $A_0$. This is valid 
for our two models with coprime $p,q$, otherwise additional terms appear. Note that this is a remarkably simple form, given the usual difficulty in computing GBZ\cite{Yokomizo2019,Yang2020a}.
We obtain a circular GBZ $\beta(\theta)=|\beta|\ee^{\ii\theta}$ of radius $|\beta|=\sqrt[2q]{A_{-}/A_{+}}$, where, for the two-band model, $A_\pm = A_\pm(\tilde{k}_x, \gamma_y, \kappa-\ii\gamma_x,  p, q)$.
For the one-band model, it is simply $|\beta| = \sqrt{(1-\gamma_y)/(1+\gamma_y)}$, which is \textit{independent} of $q$, $p$ and $\tilde{k}_x$. 
This indicates two important features -- the magnetic suppression is (i) in general \textit{not} necessarily related to the shape of GBZ, in contrast to the naive expectation that magnetic suppression deforms GBZ towards the conventional Brillouin zone; (ii) a new nonperturbative effect different from the critical skin effect where a tiny parameter change qualitatively alters the GBZ \cite{Okuma2019,Li2020}. 
Given that the $2q$ complex $\beta$'s, which constitute the general solution and are obtained from the eigenequation, have identical modulus, this is presumably due to a subtle phase interference effect brought about by the magnetic field, whereby the skin behavior is dictated by the low-energy theory to be completely altered \cite{Yokomizo2019}. Indeed, approaching the Landau limit, a NH \textit{discrete} spectrum becomes the more appropriate description than the GBZ formalism derived by assuming continuum band structure \cite{Yokomizo2019,Kawabata2020}.

\begin{figure}[t]
\includegraphics[width=8.8cm]{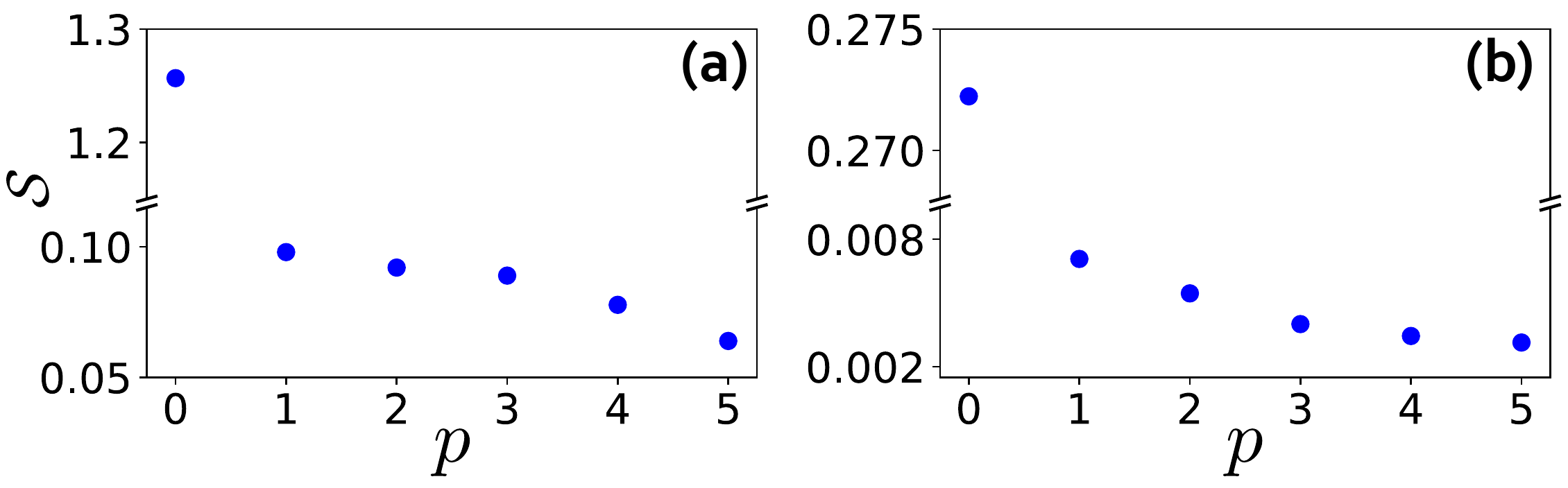}
\caption{Skin topological area $\mathcal{S}$ as a function of rational magnetic flux $2\pi p/q$ with $q=67$, $\gamma_y=0.1$, $t_x=t_y=1$ and $\gamma_x=0,\kappa=2.5$. $p=1$ approximately corresponds to the largest magnetic flux in Fig.~\ref{Fig:Landau_skin}(c,d). (a) One-band model. (b) Two-band model.}\label{Fig:Hofstadter_skin}
\end{figure}

We now turn to the skin topological area of the rational-flux models. Eq.~\eqref{eq:Det} with $\beta\rightarrow\ee^{\ii k_y}$ gives % can be rewritten as $\det[ H-E ]= A_0 + (A_+ +A_{-}) \cos(qk_y)/2 + \mathrm{i}(A_+-A_-) \sin(qk_y)/2$, which forms 
a complex-plane \textit{ellipse} with semi-axes $(A_+\pm A_-)/2$, which winds its center $A_0(E_0)$ $q$ times as $k_y$ extends from $0$ to $2\pi$. Therefore, given $E_0$, as long as zero energy falls inside the ellipse, the winding number Eq.~\eqref{eq: winding_num} is $q$, otherwise it vanishes. The skin topological area $\mathcal{S}_2(\tilde{k}_x)$ is nothing but $q$ times the region of $E_0$ such that $A_0(E_0)$ does not translate the ellipse beyond zero energy. Compared with the previous calculation that requires solving a $k_y$-dependent large open system, the relation in the thermodynamic limit is $\mathcal{S}=\lim_{N\rightarrow\infty}\frac{1}{N}\mathcal{S}_1=\frac{1}{2\pi}\int \mathcal{S}_2(\tilde{k}_x)\dd \tilde{k}_x$ where $\mathcal{S}_2(\tilde{k}_x)$ varies weakly with $\tilde{k}_x$ at large $q$. The latter reduces computational complexity and provides a simple geometric interpretation of the origin of the topological skin phenomenon. 
In Fig.~\ref{Fig:Hofstadter_skin}, we plot $\mathcal{S}$ for $q=67$ and $p=0\mathrm{-}5$ using similar parameters as in Fig.~\ref{Fig:Landau_skin}. This confirms the hypothesis of shrinking skin topological area even at larger magnetic fields beyond the low-field regime of Fig.~\ref{Fig:Landau_skin}. In particular, the jump between $p=0$ and 1, corresponding to the total change in Fig.~\ref{Fig:Landau_skin}(c,d), accurately demonstrates the magnetic suppression phenomenon by working in the thermodynamic limit naturally accessible in this approach (see SM for the analysis of finite-size effects). In addition, the skin topological area, now with uniform winding $w(E_0)\equiv q$, turns out to correspond to the brightest core regions in the previous open-lattice calculation, e.g., the brightest central core and the two side bars in Fig.~\ref{Fig:Landau_skin}(a3,b3), respectively.

\mytitle{Summary.}-- 
We find that NHSE can be strongly suppressed by an external magnetic field. The physical mechanism is closely related to that magnetic field nonperturbatively restores the low-energy description otherwise made inapplicable by NHSE. Correspondingly, we identified the topological origin of the phenomenon whereby the skin topological area, introduced in this work, shrinks as magnetic field is increased. This is observed in both irrational and rational flux models, the latter of which also demonstrates an intriguing irrelevance of standard non-Bloch band theory.

\let\oldaddcontentsline\addcontentsline% Store \addcontentsline
\renewcommand{\addcontentsline}[3]{}% Make \addcontentsline a no-op

\begin{acknowledgments}
M.L. was supported by the National Basic Research Program of China (2015CB921102) and the Strategic Priority Research Program of Chinese Academy of Sciences (XDB28000000). X.-X.Z. \& M.F. thank the Max Planck-UBC-UTokyo Center for Quantum Materials for 
financial support and were also supported by NSERC and CFREF. X.-X.Z. was partially supported by Riken Special Postdoctoral Researcher Program. 
\end{acknowledgments}\mycomment{\Yinyang}

%\nocite{*}

\bibliography{reference.bib}  % The references (bibliography) information are stored in the file named "Bibliography.bib"
\let\addcontentsline\oldaddcontentsline% Restore \addcontentsline

\newpage
\onecolumngrid
\newpage
{
	\center \bf \large 
	Supplemental Material\\
	%\large for ``Non-Hermitian exceptional Landau quantization in electric circuits"\vspace*{0.1cm}\\ 
	\large for ``\newtitle"\vspace*{0.1cm}\\ 
	\vspace*{0.5cm}
	%\newauthor
}
% \begin{center}
%     %\getauthor \\
% 	Xiao-Xiao Zhang and Marcel Franz\\
% 	\vspace*{0.15cm}
% 	\small{\textit{Department of Physics and Astronomy \& Stewart Blusson Quantum Matter Institute, University of British Columbia, Vancouver, BC, V6T 1Z4 Canada}}\\
% 	\vspace*{0.25cm}	
% \end{center}

%\twocolumngrid	

\tableofcontents

% %\clearpage
% %\appendix
% \setcounter{equation}{0}
% \setcounter{figure}{0}
% \setcounter{table}{0}
% \setcounter{page}{1}
% %\renewcommand{\theequation}{S\arabic{equation}}
% \renewcommand{\thefigure}{S\arabic{figure}}
% \renewcommand{\bibnumfmt}[1]{[S#1]}
% %\renewcommand{\citenumfont}[1]{S#1}

%%%%%%%%%% Merge with supplemental materials %%%%%%%%%%
%%%%%%%%% Prefix a "S" to all equations, figures, tables and reset the counter %%%%%%%%%%
%\appendix
\setcounter{equation}{0}
\setcounter{figure}{0}
\setcounter{table}{0}
\setcounter{page}{1}
%\makeatletter
\renewcommand{\theequation}{S\arabic{equation}}
\renewcommand{\thefigure}{S\arabic{figure}}
\renewcommand{\theHtable}{Supplement.\thetable}
\renewcommand{\theHfigure}{Supplement.\thefigure}
\renewcommand{\bibnumfmt}[1]{[S#1]}
\renewcommand{\citenumfont}[1]{S#1}
%%%%%%%%% Prefix a "S" to all equations, figures, tables and reset the counter %%%%%%%%%%

\section{Three types of low-energy effective models}

\subsection{Dirac type}\label{Sec:Dirac}
%\subsubsection{$B\neq0$}\label{Sec:Dirac_b!=0}

Let's exemplify with the $\kappa=t_y$ case of Eq.~\eqref{eq:H_2band} where the Dirac points $\bk =(\pm\frac{\pi}{2},0)$ if $\Delta=0$ or otherwise gapped by $2\Delta$.
The low-energy Hamiltonian around $\bk =(\frac{\pi}{2},0)$ is
\begin{equation}\label{eq:DiracH}
h=\sum_{i=x,y}{(t_i\Pi_i+\ii\gamma_i)\sigma_i}+\Delta\sigma_z
\end{equation}
where $\Pi_i=p_i-A_i$. Momentum $p_i$ is either taken as $-\ii\partial_i$ or $k_i$ along the open or periodic direction. In the case when $y$-direction is open, we have the eigenequation
\begin{align}\label{fff}
\left[ \sigma_x t_x(\mathfrak{K}_x+By) - \ii\sigma_y (-t_y\partial_y+\gamma_y) + \Delta\sigma_z \right]
\psi = \varepsilon \psi
\end{align}
with $\psi=(\psi_1,\psi_2)^\mathrm{T}$, which reduces to a single second-order ordinary differential equation (ODE), for instance, of $\psi_2(y)$
% \begin{equation}\label{eq:Dirac_2ndorder}
%     \psi_2'' - 2\gamma_y \psi_2' + [\tilde{\varepsilon}^2 - 2by\mathfrak{K}_x - B^2y^2] \psi_2(y)=0
% \end{equation}
% where $\tilde{\varepsilon}^2=\varepsilon^2 + \gamma_y^2 - b - \Delta^2 - \mathfrak{K}_x^2,\,\mathfrak{K}_x=k_x+\ii\gamma_x$.
\begin{equation}\label{eq:Dirac_2ndorder_shift}
    \psi_2'' - 2\tilde{\gamma}_y \psi_2' + [\tilde{\varepsilon}^2 - t_{xy}^2 B^2(y-\mathfrak{y}_0)^2] \psi_2=0
\end{equation}
where $\tilde{\varepsilon}^2=(\varepsilon^2 + \gamma_y^2 - Bt_xt_y - \Delta^2)/t_y^2,\mathfrak{y}_0=-\mathfrak{K}_x/B,\mathfrak{K}_x=k_x+\ii\gamma_x/t_x$ with $t_{xy}=t_x/t_y,\tilde{\gamma}_y=\gamma_y/t_y$. As seen here, we henceforth use fraktur letters to denote certain complex quantities for clarity and $\bk $ in low-energy models means the deviation from the band minimum.

Eq.~\eqref{eq:Dirac_2ndorder_shift} can be analytically solved in its own right as an ODE defined in the complex domain $\mathbb{C}$. 
The eigenenergy spectrum is $\varepsilon_{n>0}=\pm\sqrt{2nBt_xt_y+\Delta^2}\textrm{ and }\varepsilon_0=\Delta$, which is always real however large $\gamma_x,\gamma_y$ are, and the wavefunction component 
\begin{equation}\label{eq:phi_n}
\psi_2^{(n)}
\mycomment{=(\sqrt{\pi}l_B n! 2^{n})^{-\frac{1}{2}}  \ee^{-y^2/2l_B^2 -(k_x+\ii\gamma_x-\gamma_y)y} H_n((y+\frac{k_x+\ii\gamma_x}{b})/l_B) }
\propto \mycomment{(\sqrt{\pi}l_B n! 2^{n})^{-\frac{1}{2}}}  \ee^{-(y-\mathfrak{y}_0)^2/2l_B^2+\tilde{\gamma}_yy} H_n((y-\mathfrak{y}_0)/l_B)
\end{equation} 
where the magnetic confinement length scale $l_B=(t_{xy}B)^{-1/2}$ and $H_n$ is the $n$th Hermite polynomial holomorphic on $\mathbb{C}$. Note that this wavefunction is physically readily interpreted as an imaginary shift due to $\gamma_x$ in real-space $y$-direction as seen in $\mathfrak{y}_0$, together with another imaginary shift in momentum $p_y$ due to $\tilde{\gamma}_y$, which are fully absent in Hermitian situations. 

%\subsubsection{$B=0$}
It is crucial to note that the foregoing solution and the associated physical picture of imaginary shift in real and momentum spaces would be nullified were it not for the nonperturbative presence of an external magnetic field: Eq~\eqref{eq:phi_n} becomes unbounded or illdefined as $B\rightarrow0$. 
In fact, when $B=0$, Eq.~\eqref{eq:Dirac_2ndorder_shift} reduces to 
\begin{align}\label{eq:}
\psi_2''-2\tilde{\gamma}_y\psi_2'+(\mathcal{K}^2+\tilde{\gamma}_y^2)\psi_2=0
\end{align}
where $\mathcal{K}=(\varepsilon^2-\Delta^2-\mathfrak{K}_x^2)^{\frac{1}{2}}/t_y$.
The general solutions, $\psi_2(y)=\ee^{\tilde{\gamma}_yy}\sin{\mathcal{K}y},\ee^{\tilde{\gamma}_yy}\cos{\mathcal{K}y}$, by no means can yield a wavefunction not divergent at $y=\pm\infty$ for the presence of finite $\tilde{\gamma}_y$. The only exception is the trivial Bloch solution $\psi_2(y)\propto\ee^{\ii k_y y}$ and the complex $\varepsilon$ is specified by  $\mathrm{Im} \mathcal{K}=\pm\tilde{\gamma}_y,\mathrm{Re} \mathcal{K}=k_y$. This, however, is invalidated by the skin effect, i.e., failure of the low-energy model as discussed in the main text, unless the system has a PBC.

\subsection{Schr\"odinger type}\label{Sec:Schroedinger}
\begin{figure}[hbt]
\includegraphics[width=12.6cm]{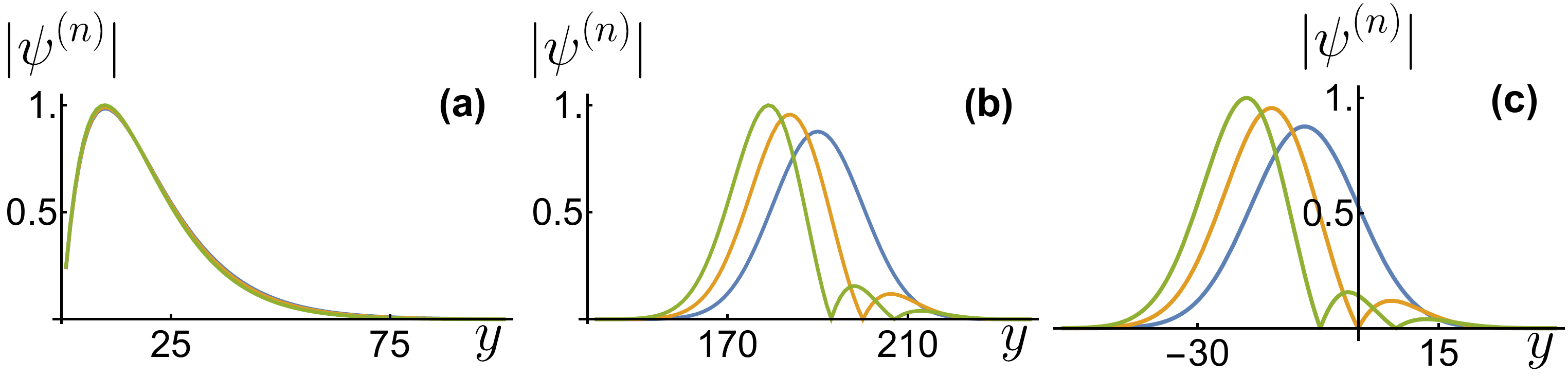}
\caption{Amplitude of three lowest energy states ($n=0$ blue, $n=1$ orange, and $n=2$ green) from the wavefunction of the Schr\"odinger-type Hamiltonian. Spatial coordinate $y$ is in units of lattice constant set to unity. Lattice calculations (a,b) with $N=400$ sites along $y$-direction take $y=0$ as the left boundary: (a) skin states aggregate towards the left boundary and do not have low-energy description when magnetic field $B=0$; (b) finite $B=0.01$ moves these states back to the bulk. (c) Low-energy exact solutions defined along $y\in(-\infty,\infty)$ are recovered by magnetic field and their profiles match with the lattice result (b). Parameters $\gamma_y=0.1,t_x=t_y=1$.}\label{Fig:wave_functions_a}
\end{figure}

%\warn{Here all parameters without factor 2 in Eq.~\eqref{eq:H_1band1}.}
The low-energy model of Eq.~\eqref{eq:H_1band1} around $\bk =(\pi,\pi)$ is
\begin{equation}\label{eq:SchroedingerH}
    h = -m + t_x k_x^2  + t_y (k_y - \ii\frac{\gamma_y}{t_y})^2 
\end{equation}
where $m=2t_x+2t_y-\frac{\gamma_y^2}{t_y}$. When $y$-direction is open, we have $k_x\rightarrow k_x-A_x$ and the eigenequation
% \begin{equation}
%     h = -m + \frac{1}{2}t_x (k_x-A_x)^2  + \frac{1}{2}t_y  (k_y - \ii\frac{\gamma}{t_y})^2 
% \end{equation}
\begin{equation}
    [-m + t_x (k_x+By)^2  - t_y  (\partial_y + \frac{\gamma_y}{t_y})^2 ]\phi(y) = \varepsilon \phi(y),
\end{equation}
which is in the \textit{same} form as Eq.~\eqref{eq:Dirac_2ndorder_shift}  with $\gamma_x=0,l_B=(t_{xy}B^2)^{-\frac{1}{4}},\varepsilon=(2n+1)\sqrt{B^2t_xt_y}-m$ for $n\in\mathbb{Z}$. It thus shares similar physical features. This is not surprising since Eq.~\eqref{eq:SchroedingerH} can be seen as a nonrelativistic Schr\"odinger version of Eq.~\eqref{eq:DiracH}: an apparently imaginary shift of momentum $k_y$ is introduced to a quasiparticle with quadratic dispersion in Eq.~\eqref{eq:SchroedingerH}; Eq.~\eqref{eq:Dirac_2ndorder_shift} combines the two linear dispersion components to effectively get a quadratic one. Also note that due to the absence of $\gamma_x$, the wavefunction will become slanting but still nodeful. This is shown in Fig.~\ref{Fig:wave_functions_a}.

On the other hand, when $x$-direction is open, we have 
% \begin{equation}
%     h = -m + \frac{1}{2}t_x (k_x)^2  + \frac{1}{2}t_y  (k_y - A_y - \ii\frac{\gamma}{t_y})^2 
% \end{equation}
\begin{equation}
    [-m - t_x \partial_x^2  + t_y  (k_y - Bx - \ii\frac{\gamma_y}{t_y})^2 ]\phi(y) = \varepsilon \phi(y)
\end{equation}
which is again in the form of Eq.~\eqref{eq:Dirac_2ndorder_shift}. 

\subsection{Mixture type}
\begin{figure}[hbt]
\includegraphics[width=12.6cm]{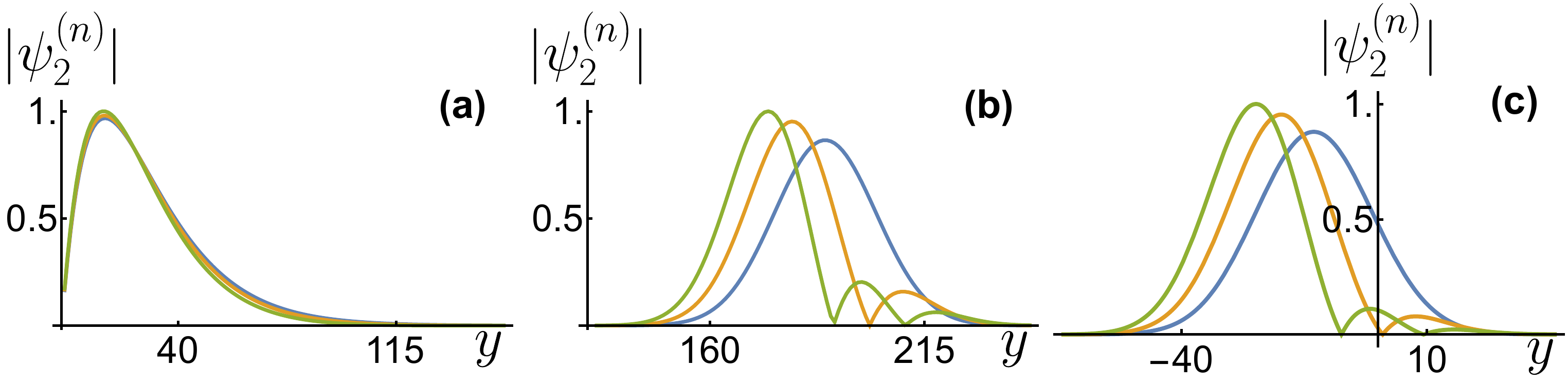}
\caption{Amplitude of three lowest energy states ($n=0$ blue, $n=1$ orange, and $n=2$ green) from the second wavefunction component $\psi_2$ of the mixed type Hamiltonian. Spatial coordinate $y$ is in units of lattice constant $a$ set to unity. Lattice calculations (a,b) with $N=400$ sites along $y$-direction take $y=0$ as the left boundary: (a) skin states aggregate towards the left boundary and do not have low-energy description when magnetic field $B=0$; (b) finite $B=0.01$ moves these states back to the bulk. (c) Low-energy exact solutions defined along $y\in(-\infty,\infty)$ are recovered by magnetic field and their profiles match with the lattice result (b). Parameters $\gamma_x=0,\gamma_y=-0.1,\kappa=2.5,t_x=t_y=1$.}\label{Fig:wave_functions_c}
\end{figure}

When $\mycomment{\kappa_1>t_x+\kappa}\kappa>t_x+t_y$ the low-energy Hamiltonian of Eq.~\eqref{eq:H_2band} around the band minimum $\bk =(\pi,0)$ is
\begin{equation}
    h=[-\mathfrak{m}-\frac{1}{2}t_xk_x^2]\sigma_x +(t_yk_y+\ii\gamma_y)\sigma_y +\Delta\sigma_z
\end{equation}
with $\mathfrak{m}=m-\ii\gamma_x,m=\kappa-t_y-t_x$. 
When $y$-direction is open, similar to Sec.~\ref{Sec:Dirac}, we have $k_x\rightarrow k_x-A_x$ and the eigenequation
\begin{equation}\label{eq:mix_type}
    a^2\psi_2'' - 2a\tilde{\gamma}_y \psi_2' + [\tilde{\varepsilon}^2 + a\tilde{B}^2 \mathsf{y} - (\tilde{\mathfrak{m}} + \tilde{B}^2\mathsf{y}^2/2)^2 ] \psi_2(\mathsf{y})=0
\end{equation}
where we make the coordinate shift $y=\mathsf{y}-k_x/B$ and define $\tilde{\varepsilon}^2=(\varepsilon^2-\Delta^2)/t_y^2, t_{xy}=t_x/t_y,\tilde{\gamma}_y=\gamma_y/t_y,\tilde{\mathfrak{m}}=\mathfrak{m}/t_y, \tilde{B}=at_{xy}^\frac{1}{2}B$. 
% \begin{equation}
%     \psi_2'' - 2\gamma_y \psi_2' + [\varepsilon^2 + \gamma_y^2 - (\frac{b^2y^2}{2}+\mathfrak{m})^2 + b^2y] \psi_2(y)=0
% \end{equation}
This completely goes beyond the conventional relativistic or nonrelativistic Landau level and even the NH Landau level in Sec.~\ref{Sec:Dirac} and Sec.~\ref{Sec:Schroedinger}.
Note that here we recover the appearance of lattice constant $a$ for the inhomogeneity between the dispersions along $x$ and $y$, which will eventually give rise to an $a$-dependence in some characteristic quantity, in contrast to the homogeneous Schr\"odinger and Dirac cases.
In fact, substituting $\psi_2(\mathsf{y})=\ee^{[\mp \mathsf{y}(\frac{a\mathsf{y}^2}{2r^3}+\tilde{\mathfrak{m}})+\tilde\gamma_y\mathsf{y}]/a}u(z_\pm)$ with $\mathsf{y}=\pm r z_\pm$ where we denote $r=(\frac{3a}{\tilde{B}^2})^{\frac{1}{3}}$, it reduces to the \textit{triconfluent Heun equation}\cite{Ronveaux1995}
\begin{equation}\label{eq:HeunT}
    u''(z)-(3z^2+\gamma)u'(z)+(\alpha-(3-\beta) z)u(z)=0
\end{equation}
where $\tilde{r}=r/a,\alpha=\tilde{r}^2(\tilde\varepsilon^2- \tilde{\gamma}_y^2),\beta=\beta_\pm=\pm3,\gamma=2\tilde{r}\tilde{\mathfrak{m}},z=z_\pm\mycomment{=\pm (\frac{3}{\tilde{b}^2})^{-\frac{1}{3}} \mathsf{y}}$. Therefore, we arrive at the general solution
\begin{equation}\label{eq:HeunGenSol}
    \psi_2(\mathsf{y})= \sum_{s=\pm} C_s\, \mycomment{\ee^{(\tilde\gamma_y-s\tilde{\mathfrak{m}})\mathsf{y} -  \frac{z_s^3}{2}}}
    \ee^{- (z_s^3+\gamma z_s)/2 + s\tilde{\gamma}_y\tilde{r}z_s}
    \mathscr{H}_\mathrm{T}(\alpha,\beta_s,\gamma,z_s)
\end{equation}
with $C_\pm$ the integration constants and $\mathscr{H}_\mathrm{T}$ the triconfluent Heun function. This can be seen from evaluating the Wronskian at any particular point, which is justified by the Abel's identity\cite{WilliamE.Boyce2012,Teschl2014}. %One can use Frobenius method to have $\mathscr{H}_\mathrm{T}$ in the form of an infinite series. 
Note that for our case $\beta\neq3(n+1),n\in\mathds{Z}_+$, $\mathscr{H}_\mathrm{T}$ does not truncate into the polynomial solution subspace because Eq.~\eqref{eq:HeunT} has an \textit{irregular} singularity at $z=\infty$\cite{Ronveaux1995,Slavyanov2000}. %the asymptotics is not solely given by the exponential factor in Eq.~\eqref{eq:HeunGenSol}. 
We can impose the Dirichlet boundary condition at $\pm\infty$ and obtain
the eigenenergy in $\alpha$ and the coefficient $C_-$ for the wavefunction when we set $C_+=1$ without loss of generality, which can be calculated with root-finding methods. We show in Fig.~\ref{Fig:wave_functions_c} the representative wavefunctions. The exponential factor in Eq.~\eqref{eq:HeunGenSol} suggests an unconventional magnetic length scale $l_B=(\frac{6}{at_{xy}B^2})^{\frac{1}{3}}$, different from the Schr\"odinger and Dirac cases. To assure this, one can adopt a WKB (Wentzel–Kramers–Brillouin) ansatz $\psi_2(\mathsf{y})\sim\ee^{f(\mathsf{y})}$ and retain the leading order terms, which gives rise to $a^2f'(\mathsf{y})^2=(\tilde{B}^2\mathsf{y}^2/2)^2$ and hence $f(\mathsf{y})=-\frac{at_{xy}B^2}{6}|\mathsf{y}|^3$.

This magnetic suppression of skin effect originates physically from the similar picture as the previous two cases, except that the quadratic and in general \textit{complex} confinement potential is replaced by an even more involved quartic one. Here, although lacking an analytically transparent form, the wavefunction actually inherits the \textit{same} key features as Eq.~\eqref{eq:phi_n}: eigenstate labeled by integer $n$ bears $n$ nodes and $n+1$ peaks, which is not automatically guaranteed for this singular Sturm-Liouville problem\cite{WilliamE.Boyce2012,Teschl2014}; $\gamma_x,\gamma_y$ respectively removes nodes and slants amplitudes along the space. %Note that the term linear in $y$ in Eq.~\eqref{eq:mix_type} introduces a magnetic field-dependent spatial asymmetry even in the $\gamma_y=0$ case. 

Lastly, we point out a distinct feature from previous two types. The corresponding ODE of $\psi_1$ only differs from Eq.~\eqref{eq:mix_type} in the sign of the linear-$\mathsf{y}$ term, which therefore shares the \textit{same} quantum number. Consequently, two components of $\psi=(\psi_1,\psi_2)$ share the similar profile of peaks and (slightly offset) nodes. Actually, only a minor difference is brought by this sign switch.

On the other hand, when $x$-direction is open, we have
\begin{equation}
    \psi_2^{(4)} - 4\tilde{\mathfrak{m}}\psi_2'' + 4\ii t_{yx} B \psi_2' - 4 [\tilde{\varepsilon}^2-B^2t_{yx}^2(x-\mathfrak{x}_0)^2]\psi_2 = 0
\end{equation}
where $\mathfrak{x}_0=(k_y+\ii\gamma_y/t_y)/B,\tilde{\mathfrak{m}}=\mathfrak{m}/t_x,\mathfrak{m}=m-\ii\gamma_x,m=\kappa-t_y-t_x,t_{yx}=t_y/t_x,\tilde{\varepsilon}^2=(\varepsilon^2-\Delta^2-\mathfrak{m}^2)/t_x^2$. 
Since this 4th-order ODE is not likely to be analytically accessible, we can numerically solve it and again obtain well-defined bulk solutions.

%\subsection{Summary}

\section{Finite-size effect in the magnetic suppression}
Here, as a complement to Fig.~\ref{Fig:Landau_skin}(c,d) in the main text, we present additional data of different system sizes up to the computational ability we have at hand. Fig.~\ref{Fig:Landau_skin} in the main text uses the largest common system size ($N=200$) we can reach for both one-band and two-band models. Note that this calculation of the skin topological area is considerably more demanding than wavefunction calculations such as those for Fig.~\ref{Fig:wave_functions}.

\begin{figure}[hbt]
\includegraphics[width=12.6cm]{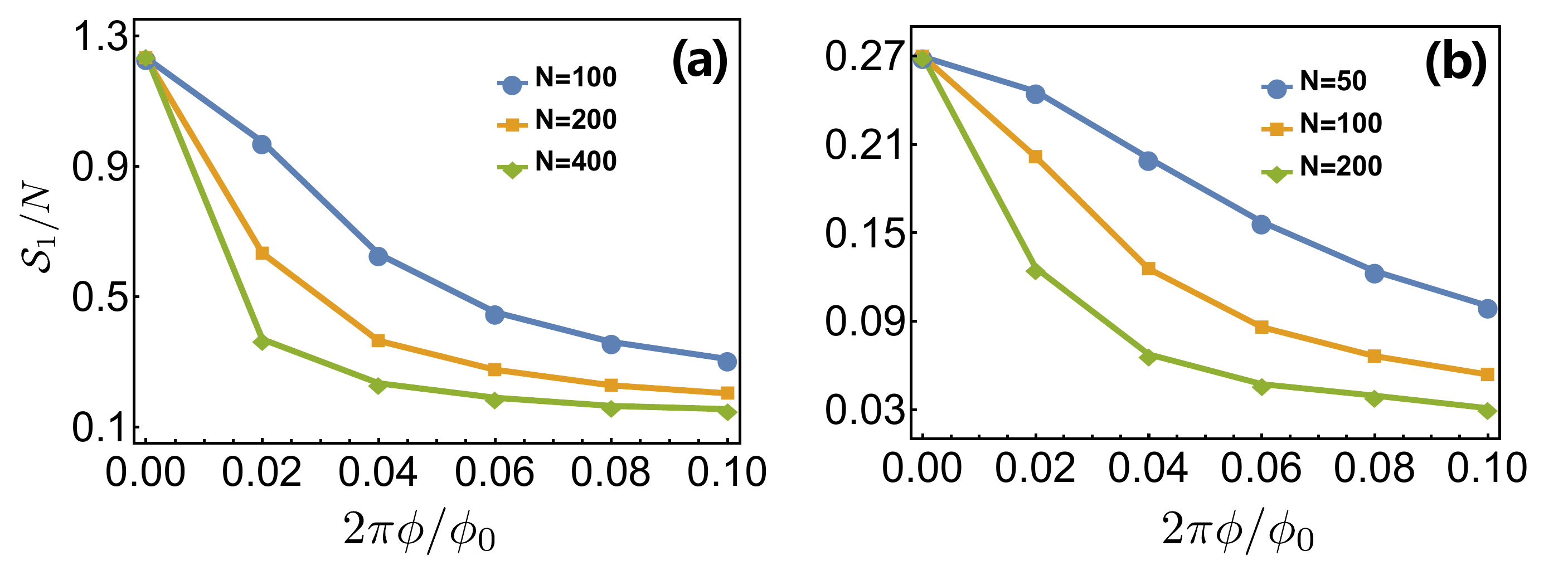}
\caption{Skin topological area $\mathcal{S}_1/N$ as a function of magnetic flux at different sizes $N$ of the open system. Other parameters same as Fig.~\ref{Fig:Landau_skin} in the main text, which uses the common size $N=200$. (a) One-band case: we compare $N=100,200,400$; (b) two-band case: we compare $N=50,100,200$. }\label{Fig:finitesize}
\end{figure}

In Fig.~\ref{Fig:finitesize}, the important feature is the common trend that a larger system size lowers the skin topological area at finite $B$. Note that the rightmost magnetic flux $2\pi\phi/\phi_0=0.1$ roughly corresponds to the $p=1$ point in Fig.~\ref{Fig:Hofstadter_skin} since $2\pi\cdot 1/67\approx0.1$. Consequently, the jump in the area between zero field and  $2\pi\phi/\phi_0=0.1$ increases as the system size grows, which will eventually reach the thermodynamic-limit result in Fig.~\ref{Fig:Hofstadter_skin}. For instance, the lowest area slightly above 0.1 in Fig.~\ref{Fig:finitesize}(a) is well within the same order of magnitude to the corresponding $p=1$ point about 0.1 in Fig.~\ref{Fig:Hofstadter_skin}(a), especially given that the zero-field values match between two calculations and are order-of-magnitude larger. This helps to confirm the thermodynamic-limit nature of the rational-flux calculations.

\section{Magnetic suppression of NHSE as a global property}\label{Sec:SkinArea}
\begin{figure}[hbt]
\includegraphics[width=8.6cm]{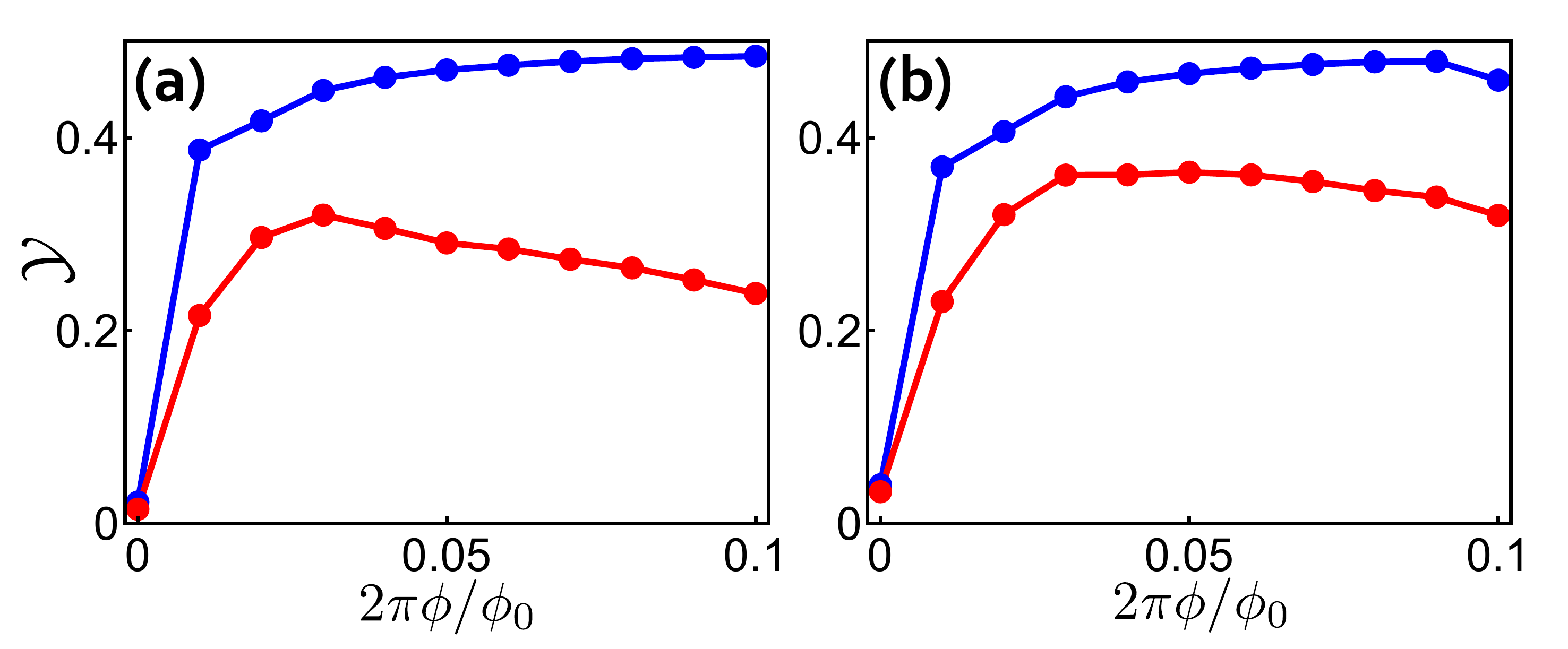}
\caption{Average center $\mathcal{Y}$ of the wavefunctions as a function of magnetic flux. Blue lines are for the lowest-energy $10\%$ eigenstates and the red lines are for all the eigenstates. Lattice site $N=400$, $t_x=t_y=1$, $\gamma_y=0.1$ and $\kappa=2.5$. (a) One band model. (b) two-band model.}\label{Fig:localization}
\end{figure}

% \subsection{Center of the wavefunctions}
To show the magnetic suppression as a global property, we define the average center of the wavefunctions as 
\begin{equation}
    \mathcal{Y}=\frac{1}{2\pi M N} \int_0^{2\pi} \mathrm{d}k_x \sum_{m=1}^{M}\sum_{j=1}^N j|\psi_{k_x,m}(j)|^2
\end{equation}
for the one-band model, where $\psi_{k_x,m}(j)$ is the wavefunction amplitude at site $j$ of the $m$-th lowest state labeled by $k_x$.  The number of the unit cells is $N$ and the sum is over the lowest $M$ eigenstates. Similar expression is applicable for the two-band model, except $N\to 2N$ due to doubled dimension. $\mathcal{Y}\in[0,1]$ by definition, and $\mathcal{Y}=0$ ($1$) corresponds to that all wavefunctions considered are localized at the left (right) edge.  As shown in Fig.~\ref{Fig:localization}, without magnetic field, the wavefunctions are localized near the left edge. When the magnetic field is turned on, the NHSE is suppressed and the average center of the wavefunctions moves towards the center. Compared with the high-energy wavefunctions, the low-energy ones are more sensitive to the magnetic field as per our physical picture from the low-energy theory, whose average center can approach $1/2$, i.e., the bulk center, with a moderate magnetic field strength. This is consistent with our analytical results of the corresponding low-energy continuum models.

One should note that this is not a complete elimination of NHSE. The discussion gains us from low-energy exact solutions an intuitive and clear physical picture of the magnetic suppression of NHSE. Although this generally can reach well beyond low-energy states as we see, one would naturally imagine extra complexities: high-energy behavior may not render all states back to the bulk or free from skin aggregation; for large magnetic fields when $l_B$ becomes comparable to the lattice constant, the transparent low-energy picture ceases to be fully accurate. Indeed, some characteristic features of NHSE, including non-Bloch bands and skin topological area, diminish but not vanish.

\section{Non-Bloch band theory with rational gauge flux}\label{Sec:nonBloch}

A closely related aspect is the non-Bloch band theory of NH systems with possible skin effects, where one defines the generalized Brillouin zone (GBZ) that yields the correct prediction of band spectrum continuum under OBC in the macroscopic limit\cite{Yao2018a,Yokomizo2019,Kawabata2020}. Any deviation of a GBZ from the conventional unit-circular BZ, $\ee^{\ii k}$ for $k\in[0,2\pi)$, in the complex plane would imply the presence of skin modes. Unfortunately, this is not directly an approach amenable to adding a magnetic field, simply because any foregoing gauge realization would open at least one direction and make the system effectively macroscopically many bands, which obstructs any accessible analysis of the GBZ.

With the non-Bloch momentum substitution $\ee^{\ii k_y}\rightarrow\beta$, we find the key quantity in the general form
% \begin{equation}\label{eq:Det}
% \det[ H-E ]= f + A \cos(qk_y) + B \sin(qk_y)
% \end{equation}
\begin{equation}\label{eq:Det_1}
\det[ H-E ]= A_0 + A_+ \beta^q + A_- \beta^{-q}
\end{equation}
when $p$ and $q$ are coprime, otherwise it adds more complexity due to additional harmonics.
For model Eq.~\eqref{eq:H_1band1}, %$A(\gamma,q)=(1+\lambda)^q + (1-\lambda)^q,B(\gamma,q)=\ii[(1+\lambda)^q - (1-\lambda)^q],f(\tilde k_x,\gamma,p,q,E)=2\cos(q\tilde{k}_x) + P(\gamma,p,q,E^{(q)})$
$A_\pm(\gamma_y,q)=(1\pm\gamma_y)^q,A_0(\tilde k_x,\gamma_y,p,q,E)=2\cos(q\tilde{k}_x) + P(\gamma_y,p,q,E^{(q)})$
% \begin{align*}
%     \det \left[ H(\tilde{k}_x, k_y)-E_0 \right] =[(1+\lambda)^q + (1-\lambda)^q]\cos (qk_y)\\
%   +\mathrm{i} [(1+\lambda)^q - (1-\lambda)^q]\sin (qk_y) + 2\cos(q\tilde{k}_x) + P_q(\gamma,p,E_0)
% \end{align*}
with $P(E^{(q)})$ a $q$-th order polynomial of $E$.
%Model Eq.~\eqref{eq:H_2band}, when $tx=ty=\kappa=1$ \warn{more general form TBD}, has more complex \warn{is $q=1$ valid?} $A(\tilde{k}_x,\gamma_y,q)=\frac{1}{2^{q-2}}\cos(q\tilde{k}_x)+P_A(\gamma_y^{(q-1)}, \kappa_1^{(q)}),B(\gamma_y,q)=\mathrm{i} P_B(\gamma_y^{(q)}, \kappa_1^{(q-1)}),f(\tilde k_x,\gamma_y,p,q,E)$, and $P_{A,B}(\gamma_y^{(m)}, \kappa_1^{(n)})$ stands for some polynomial of degree $m$ and $n$ respectively in $\gamma_y$ and $\kappa_1$.
Model Eq.~\eqref{eq:H_2band}, for instance, when $t_x=t_y=1,\gamma_x=0$, bears a more complex
form $A_\pm(\tilde{k}_x,\gamma_y,\kappa,p,q)=\frac{1}{2^{q-1}}\left[\cos(q\tilde{k}_x)+P((\mp\gamma_y)^{(q)}, \kappa^{(q)})\right]$ and $A_0=A_0(\tilde k_x,\gamma_y,\kappa, p,q,E)$ where $P(x^{(m)}, y^{(n)})$ is a polynomial of degree $m$ and $n$ respectively in $x$ and $y$.

Solving the eigenequation $\det[ H(\beta)-E ]=0$, we obtain $2q$ solutions dependent on $E$: $|\beta_1|\leq|\beta_2|\leq\cdots\leq|\beta_{2q-1}|\leq|\beta_{2q}|$. GBZ is the trajectory specified by $|\beta_q|=|\beta_{q+1}|$ \cite{Yokomizo2019}. Since the non-Bloch spectrum $E$ is not known \textit{a priori}, we practically subtract $\det[ H(\beta)-E ]=0$ by its counterpart $\det[ H(\beta\ee^{\ii\theta})-E ]=0$ with a distinct complex momentum of identical modulus
\begin{equation}\label{eq:solveGBZ}
    \det[ H(\beta)-E ]-\det[ H(\beta\ee^{\ii\theta})-E ]=0
\end{equation}
and solve for the $q$th larger $\beta(\theta)$ in absolute value. GBZ is given by $\beta(\theta),\theta\in[0,2\pi)$. This procedure is most convenient when the left-hand side of Eq.~\eqref{eq:solveGBZ} no longer depends on $E$, which is indeed the case for our models as $E$ solely enters $A_0$ in Eq.~\eqref{eq:Det}.

For model Eq.~\eqref{eq:H_1band1}, GBZ is simply a non-unit circle $|\beta| = \sqrt{(1-\gamma_y)/(1+\gamma_y)}$, which is \textit{independent} of $q$, $p$ and $\tilde{k}_x$. This undoubtedly exemplifies a significant conclusion -- the magnetic suppression of skin effect is in general \textit{not} necessarily related to the shape of GBZ, in contrast to what one might naively expect, i.e., magnetic suppression deforms the GBZ towards the conventional BZ. Without the magnetic field, the GBZ of the two-band model is $|\beta|=\sqrt{|F_0(\gamma_y)/F_0(-\gamma_y)|}$, where $F_0(\gamma_y)=\kappa+t_x\cos k_x+\gamma_y$, where the GBZ may collapse or expand to infinity for some particular $k_x$ in some cases. With the rational magnetic flux, $|\beta|=\sqrt[2q]{A_{-}/A_{+}}$, where $A_\pm = A_\pm(\tilde{k}_x, \gamma_y, \kappa, p, q)$ as defined above.

\begin{figure}[hbt]
\includegraphics[width=12.6cm]{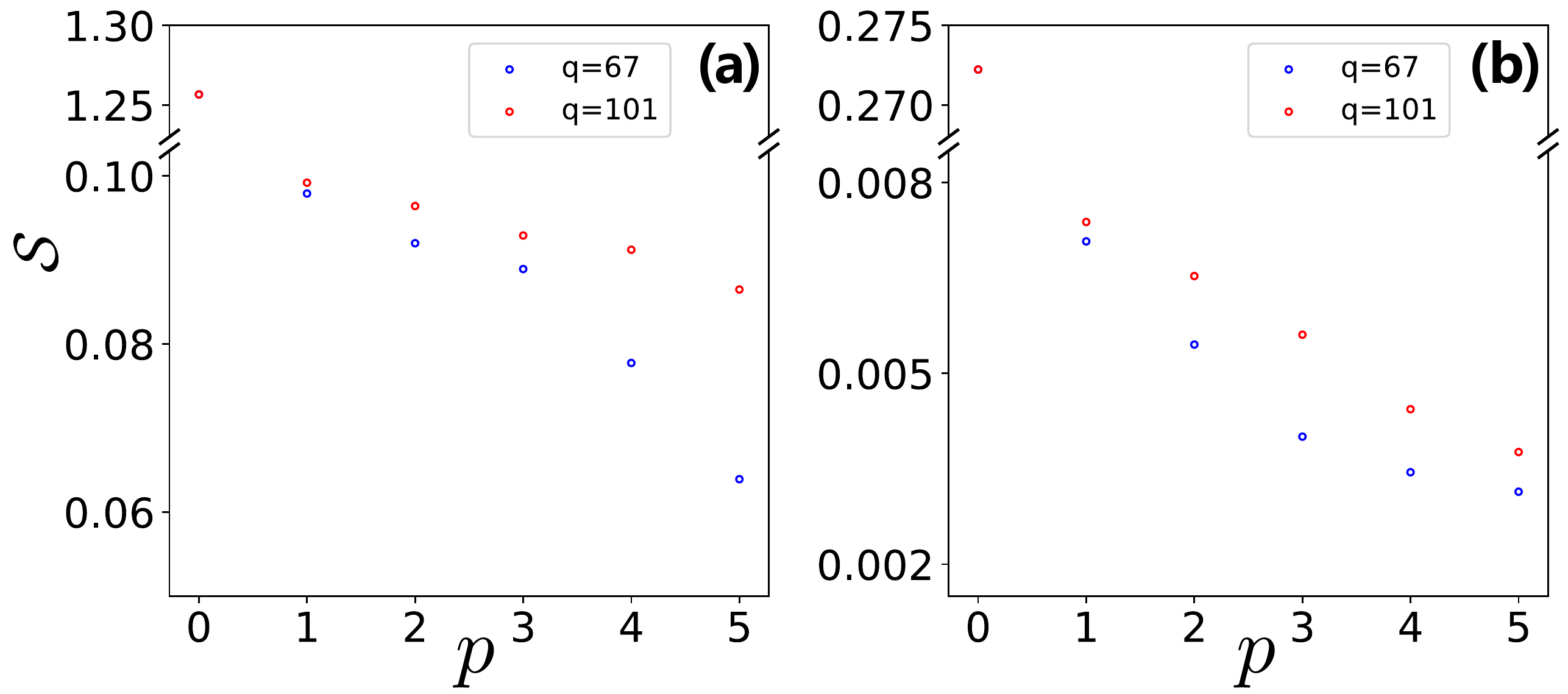}
\caption{Skin topological area $\mathcal{S}$ as a function of rational magnetic flux $2\pi p/q$ with $q=67,101$.
Other parameters same as Fig.~\ref{Fig:Hofstadter_skin} in the main text. (a) One-band case; (b) two-band case.}\label{Fig:q-dependence}
\end{figure}

Lastly, we present the additional data on the $q$-dependence of the suppression effect in Fig.~\ref{Fig:q-dependence} as smaller $q$ signifies larger magnetic field. The clear overall feature is that larger $q$ in general leads to smaller suppression effect, as one would presume on the physical ground. 
Note that in those rational-flux calculations the magnetic field is in general large and beyond the low-field limit where the low-energy theory applies the best. It is thus remarkable that the growing skin suppression effect with magnetic field can hold to a large extent, especially given that rational-flux models can typically introduce accidental features dependent on the particular rational fraction. Note also that the $p=0$ points coincide for different $q$, because the system physically remains to be the same one without magnetic field regardless of the choice of the enlarged unit cell.

\section{Hamiltonians with magnetic fluxes}
\subsection{Lattice model open in \texorpdfstring{$y$}{y}-direction}
To show NHSE and the magnetic suppression from the wavefunctions, we need to open $y$-direction and choose the gauge $\bA =B(-y,0)$. The $x$-direction can be treated with the PBC. The Hamiltonians read as:
\begin{equation}\label{eq:finite-size-openY-1band}
    H(k_x)=\sum_{n=1}^N \left( 2t_x\cos(k_x-2\pi n\phi/\phi_0) c_n^{\dagger}c_n + (t_y-\gamma_y) c_{n+1}^{\dagger} c_n + (t_y+\gamma_y) c_n^{\dagger}c_{n+1} \right)
\end{equation}
for one-band model and 
\begin{align}\label{eq:finite-size-openY-2band}
    H(k_x)&=\sum_{n=1}^N \bigg[ \Delta (a_n^{\dagger}a_n-b_n^{\dagger}b_n) + \left(-\kappa+\gamma_y-t_x\cos(k_x-2\pi n\phi/\phi_0)\right) a_{n}^{\dagger} b_n +\left(-\kappa-\gamma_y-t_x\cos(k_x-2\pi n\phi/\phi_0)\right) b_{n}^{\dagger} a_n \nonumber \\    
    &\qquad+t_y(a_{n+1}^{\dagger} b_n+b_{n}^{\dagger} a_{n+1})\bigg]
\end{align}
for two-band model, where $\phi$ is the magnetic flux per unit cell and $\phi_0=h/e$ is the flux quantum.

\subsection{Lattice model open in \texorpdfstring{$x$}{x}-direction}
In order to calculate the topological winding number, we need to integrate $k_y$. Therefore, the gauge choice has to be $\bA =B(0,x)$ and the $x$ direction is open, the effective one dimensional lattice Hamiltonian writes: 
\begin{equation}\label{eq:finite-size-openX-1band}
\begin{split}
    H(k_y)=  \sum_{n=1}^{N-1} t_x c_n^\dagger c_{n+1} +\mathrm{h.c.} + \sum_{n=1}^N \left[2t_y \cos(k_y+2\pi n\phi/\phi_0)+2\mathrm{i}\gamma_y\right] c_n^\dagger c_n
\end{split}
\end{equation}
for the one-band model and 
\begin{equation}\label{eq:finite-size-openX-2band}
    \begin{split}
&H(\tilde{k}_x,k_y)=\\
    & \sum_{n=1}^N\{-\kappa + t_y \cos(k_y-2\pi n\phi/\phi_0)-\mathrm{i}[\mathrm{i} \gamma_y+t_y\sin(k_y-2\pi n\phi/\phi_0)]\}a_n^\dagger b_n \\
    & +\sum_{n=1}^N\{-\kappa + t_y \cos(k_y-2\pi n\phi/\phi_0)+\mathrm{i}[\mathrm{i} \gamma_y+t_y\sin(k_y-2\pi n\phi/\phi_0)]\}b_n^\dagger a_n \\
    &+\frac{t_x}{2} \sum_{n=1}^{N-1} (a_n^\dagger b_{n+1}+ b_{n+1}^\dagger a_{n} + b_n^\dagger a_{n+1}+a_{n+1}^\dagger b_{n})
    \end{split}
\end{equation}
for the two-band model, where $k_y\in [0,2\pi)$.

\subsection{Hamiltonians with a rational magnetic flux}
With the gauge choice $\bA =B(0, x)$ and a rational gauge flux $p\phi_0/q$ per unit cell, the $x$ direction recovers the $q$-unit cell translational symmetry. The Hamiltonians write as,
\begin{equation}\label{eq:Hofstadter-1band}
\begin{split}
    &H(\tilde{k}_x,k_y)=  \sum_{j=1}^q t_x e^{\mathrm{i}\tilde{k}_x} c_j^\dagger c_{j+1} +\mathrm{h.c.} + \sum_{j=1}^q \left[2t_y \cos(k_y+2\pi pj/q)+2\mathrm{i}\gamma_y\right] c_j^\dagger c_j
\end{split}
\end{equation}
for the one-band model and 
\begin{equation}\label{eq:Hofstadter-2band}
    \begin{split}
&H(\tilde{k}_x,k_y)=\\
    & \sum_{j=1}^q\{-\kappa + t_y \cos(k_y-2\pi pj/q)-\mathrm{i}[\mathrm{i} \gamma_y+t_y\sin(k_y-2\pi pj/q)]\}a_j^\dagger b_j \\
    & +\sum_{j=1}^q\{-\kappa + t_y \cos(k_y-2\pi pj/q)+\mathrm{i}[\mathrm{i} \gamma_y+t_y\sin(k_y-2\pi pj/q)]\}b_j^\dagger a_j \\
    &+\frac{t_x}{2} \sum_{j=1}^q (e^{\mathrm{i}\tilde{k}_x} a_j^\dagger b_{j+1}+e^{-\mathrm{i}\tilde{k}_x} b_{j+1}^\dagger a_{j} + e^{\mathrm{i}\tilde{k}_x} b_j^\dagger a_{j+1}+e^{-\mathrm{i}\tilde{k}_x} a_{j+1}^\dagger b_{j})
    \end{split}
\end{equation}
for the two-band model, with $\tilde{k}_x \in [0,2\pi/q),k_y\in [0,2\pi)$. Note also that the PBC identifies $a_{q+1}\equiv a_1$ and $b_{q+1}\equiv b_1$.

\end{document}